# Surface vortex solitons

Yaroslav V. Kartashov,[1]* Alexey A. Egorov,[2] Victor A. Vysloukh,[3] and Lluis Torner[1]

[1]*ICFO-Institut de Ciencies Fotoniques, Mediterranean Technology Park, 08860 Castelldefels (Barcelona), Spain*

[2]*Physics Department, M. V. Lomonosov Moscow State University, 119899, Moscow, Russia*

[3]*Departamento de Fisica y Matematicas, Universidad de las Americas – Puebla, Santa Catarina Martir, CP 72820, Puebla, Cholula, Mexico*

We discover the existence of vortex solitons supported by the surface between two optical lattices imprinted in Kerr-type nonlinear media. Such solitons can feature strongly noncanonical profiles, and we found that their properties are dictated by the location of the vortex core relative to the surface. The refractive index modulation forming the lattices at both sides of the interface results in complete stability of the vortex solitons in wide domains of their existence, thus introducing the first known example of stable topological solitons supported by a surface.

Vortex solitons, i.e. localized excitations which carry screw topological phase dislocations in nonlinear materials play a central role in many branches of physics, including superfluids, plasmas, Bose-Einstein condensates, and nonlinear optics (for recent reviews, see, e.g., [1]). Bright vortex solitons are typically prone to azimuthal modulation instabilities that cause their self-splitting into ground-state solitons that fly off the original state. Only a few examples of stable vortices in uniform materials were found to date, including vortices in media with competing nonlinearities, dissipative systems, and, very recently, in nonlocal media [2]. Spatial modulation of parameters of nonlinear media can have strong stabilizing action on vortices. It was shown in Ref. [3] that stable vortices can exist in discrete lattices, e.g. two-dimensional arrays of weakly coupled waveguides. This proposal was further confirmed theoretically for vortices in continuous lattices imprinted in cubic and saturable nonlinear materials [4]. Lattice vortex solitons have been observed in photorefractive crystals [5], where periodic lattices were optically induced [6,7]. Stable vortex solitons were also found in higher-order spectral bands of periodic lattices [8], in three-dimensional lattices [9], in Bessel lattices [10], in materials with quadratic nonlinearities [11], and in photonic crystals [12]. Vortex solitons can be asymmetric in symmetric [13] and in anisotropic lattices [14].

A new important possibility, never addressed to date, is the existence of vortex solitons at the surface between two different materials. Surfaces can support waves confined at the very interface, and nonlinear surface waves on various interfaces were studied in solid-state physics, in nonlinear and near-field optics (see, e.g., [15-17]). Nonlinearity drastically alters refraction scenario for optical beams and results in bistability and transition between regimes of total internal reflection and complete transmission [18]. Optical surface waves have been observed in photorefractive materials with diffusion nonlinearity [19] and at the interface of uniform and layered media, including photonic crystals [20]. Arrays of waveguides made with currently available technologies allow formation of solitons at the interface between optical lattices, a landmark concept put forward recently in [21].

In this Letter we discover the existence of vortex solitons at an interface of two periodic lattices imprinted in Kerr-type focusing nonlinear media. To the best of our knowledge, the existence of surface vortices has not been reported so far, not even for interfaces of uniform materials. We found that surface vortex solitons feature strongly asymmetric shapes and noncanonical phase distributions. Importantly, we found that the lattices forming the interface results in complete stability of the vortex solitons in wide domains of their existence. Also, we reveal nontrivial relation between interface parameters and existence domains of surface vortex solitons.

We consider propagation of laser radiation at the interface of two periodic lattices imprinted in the focusing media with Kerr-type saturable nonlinearity, described by the nonlinear Schrödinger equation for dimensionless complex amplitude of the light field $q$:

$$i\frac{\partial q}{\partial \xi} = -\frac{1}{2}\left(\frac{\partial^2 q}{\partial \eta^2} + \frac{\partial^2 q}{\partial \zeta^2}\right) - \frac{q|q|^2}{1+S|q|^2} - R(\eta,\zeta)q. \qquad (1)$$

Here the transverse $\eta$ and longitudinal $\xi$ coordinates are scaled in terms of beam width and diffraction length, respectively, and $S$ is the saturation parameter. The function $R(\eta,\zeta) = \delta p H(\eta) + (p/4)[1-\cos(\Omega\eta)][1-\cos(\Omega\zeta)]$ stands for total refractive index profile, where $p$ is the depth of periodic part of the lattice, $\Omega$ is its frequency, the function $H(\eta) \equiv 0$ for $\eta \leq 0$, and $H(\eta) \equiv 1$ for $\eta > 0$, and $\delta p$ characterizes the height of the step in the constant part of refractive index. The profile of such lattice is depicted in Fig. 1(c). Further we assume that the depth of periodic refractive index modulation and the height of refractive index step at $\eta = 0$ are small compared with unperturbed refractive index of material and are of the order of nonlinear contribution due to the Kerr effect. Such refractive index landscapes can be either fabricated (e.g., by ion implantation), or they might be induced optically in the photorefractive crystals. In the latter case the periodic part of the lattice might be created by interfering four plane waves, while nonuniform incoherent background illumination of the proper side of the crystal can produce a sharp step in the refractive index at $\eta = 0$. Standard techniques [6] based on the vectorial interactions can be implemented to observe soliton propagation in optically induced refractive index profile, that can be fine-tuned by intensities of lattice-creating waves and background illumination. It should be pointed out that other types of nonlinear lattice interfaces could potentially be realized in the photorefractive crystals, e.g. by applying different voltages to the different crystal parts. Equation (1) admits several conserved quantities, including the energy flow $U = \int \int_{-\infty}^{\infty} |q|^2 \, d\eta d\zeta$.

We search for localized vortex soliton solutions of Eq. (1) in the following form $q = [w_\mathrm{r}(\eta,\zeta) + iw_\mathrm{i}(\eta,\zeta)]\exp(ib\xi)$, where functions $w_\mathrm{r}$ and $w_\mathrm{i}$ represent the real and imaginary parts of light field, respectively, and $b$ is the propagation constant. The topological winding number (or vortex charge) $m$ of complex field $q$ can be defined by the circulation of the gradient of the field phase $\arctan(w_\mathrm{i}/w_\mathrm{r})$ around the phase singularity, where field vanishes. Further we are interested in vortex soliton solutions with unit topological charges $m$. Substitution of the light field in such form into Eq. (1) yields the system

$$\frac{1}{2}\left(\frac{\partial^2}{\partial \eta^2} + \frac{\partial^2}{\partial \zeta^2}\right)w_\mathrm{r,i} - bw_\mathrm{r,i} + \frac{w_\mathrm{r,i}(w_\mathrm{r}^2 + w_\mathrm{i}^2)}{1 + S(w_\mathrm{r}^2 + w_\mathrm{i}^2)} + pRw_\mathrm{r,i} = 0, \qquad (2)$$

Upon searching for various vortex soliton profiles we solved system (2) numerically with a standard relaxation method. It is apparent that very far from interface located at $\eta = 0$ the properties of vortex solitons supported by lattice, shown in Fig. 1(c), do not differ substantially from the properties of their counterparts in perfectly periodic lattices, but situation changes dramatically as soon as soliton energy concentrates in lattice periods adjacent to the interface. In this case the internal structure is different for vortices shifted into lattice regions with lower (at $\eta < 0$) or higher ($\eta > 0$) mean refractive index values. The simplest situation occurs when vortex phase singularity is located close to the point $\eta = \zeta = 0$. Representative examples of profiles of such vortices are depicted in Figs. 1(a) and 1(b). By analogy with perfectly periodic lattice we term such vortices "off-site" since phase singularity is located between local lattice maxima. Four main intensity lobes (whose separation is minimal for off-site case) are clearly resolvable in vortex profile, but two of them located at $\eta > 0$ are smaller than those located at $\eta < 0$, so that vortex become strongly asymmetric, especially for big refractive index steps $\delta p$. The phase distributions for asymmetric surface vortices are noncanonical [22], i.e. on a ring of radius $r$ whose center coincides with phase singularity, phase does not grow linearly, but rather possesses alternating regions of slow (in the vicinity of local intensity maxima) and fast growth. While increasing $\delta p$ results in stronger asymmetry, the growth of the periodic modulation depth $p$ leads to stronger localization of vortex energy near local lattice maxima.

We found that at fixed $\delta p$ and $p$ there exist lower $b_{\text{low}}$ and upper $b_{\text{upp}}$ cutoffs for existence of off-site surface vortices. In contrast to canonical vortices in uniform media the ratio $U_{\text{r}}/U_{\text{i}}$ of quantities $U_{\text{r,i}} = \int\int_{-\infty}^{\infty} w_{\text{r,i}}^2 d\eta d\zeta$ is not constant and varies with $b$, $p$, and $\delta p$. We found that close to both lower and upper cutoffs one of the quantities $U_{\text{r,i}}$ abruptly tends to zero. The domain of existence for surface vortices is presented in Fig. 2(a). The width of existence domain in $b$ is maximal (though still limited) for $\delta p \to 0$, and *quickly shrinks* with increase of $\delta p$, so that above certain *critical* value $\delta p_{\text{cr}}$ of step in the constant part of refractive index off-site surface vortices *do not exist* at the lattice interface. For a fixed $\delta p$ the width of existence domain gets broader with decrease of the lattice depth $p$. Though dependencies $U_{\text{r,i}}(b)$ are nonmonotonic close to cutoffs, the total energy flow $U$ still is a monotonically increasing function of propagation constant (Fig. 2(b)). The maximal energy flow carried by the vortex soliton residing at the interface quickly decreases with increase of $\delta p$. We found that close to lower cutoff surface vortices are typically less localized than their counterparts near upper cutoff. This is especially pronounced at small and moderate values of $\delta p \sim 1$, when low-energy vortices expand over several neighboring lattice maximums, moreover, the expansion in the region $\eta > 0$ can be much more pronounced than that for $\eta < 0$. Vortices with strongly asymmetric shapes at $\delta p \gg 1$ are typically well localized in both cutoffs. The striking feature of surface vortices is that critical value of refractive index step $\delta p_{\text{cr}}$ *decreases* with increase of the lattice depth (Fig. 2(c)), i.e. *stronger asymmetries* can only be achieved at interfaces formed by *shallower* lattices.

One of the central results of this Letter is that strongly asymmetric vortices at interface of two lattices can be made *completely stable* in the substantial part of their existence domain. To analyze stability of surface vortex solitons, we searched for the perturbed solutions of Eq. (1) in the form $q = (w_r + iw_i + u_r + iu_i)\exp(ib\xi)$, where $u_r(\eta,\zeta,\xi)$ and $u_i(\eta,\zeta,\xi)$ are real and imaginary parts of perturbation that can grow with complex rate $\delta$ upon propagation. Linearization of Eq. (1) around $w_r, w_i$ yields the system of equations

$$\left( \frac{3w_{r,i}^2 + w_{i,r}^2 + S(w_r^2 + w_i^2)^2}{[1 + S(w_r^2 + w_i^2)]^2} u_{r,i} + \frac{2w_r w_i}{[1 + S(w_r^2 + w_i^2)]^2} u_{i,r} \right) + \\ + \frac{1}{2}\left( \frac{\partial^2}{\partial \eta^2} + \frac{\partial^2}{\partial \zeta^2} \right) u_{r,i} - bu_{r,i} + pRu_{r,i} = \pm \frac{\partial}{\partial \xi} u_{i,r} \quad (3)$$

where we suppose that $u_{r,i} \sim \exp(\delta\xi)$. We solved this system numerically in order to find the perturbation profiles and associated growth rates $\delta$. The outcome of stability analysis for off-site surface vortices performed for various sets of parameters $\delta p$ and $p$ is summarized in Fig. 3. It was found that such strongly asymmetric vortices are *stable* in the most part of their existence domain at moderate $p \sim 1$ and high depths of periodic modulation. There are two instability domains located near the lower and upper cutoffs (Fig. 3(c)). The instability domain near upper cutoff is too narrow and not even visible in the plot. The width of the lower instability domain *decreases* with growth of $\delta p$. The instabilities encountered for the asymmetric surface vortices are associated with complex growth rates and, hence, are of oscillatory type. The typical dependencies of real part of perturbation growth rate on $b$ are depicted in Figs. 3(a) and 3(b). Notice, that increase of the depth $p$ of periodic modulation results in further reduction of the widths of lower and upper instability domains.

To confirm the results of the linear stability analysis we solved Eq. (1) directly with the input conditions $q|_{\xi=0} = (w_r + iw_i)(1 + \rho)$, where $\rho(\eta,\zeta)$ is the random function with Gaussian distribution and variance $\sigma_{\text{noise}}^2$. The results obtained confirmed the conclusions drawn from the linear stability analysis in all cases. Asymmetric surface vortices belonging to stability domain retain their input structure for huge distances, far exceeding experimentally achievable crystal lengths (see Fig.

4(a)), while unstable representatives of off-site vortex families decay via progressively growing oscillations of their intensity lobes (Fig. 4(b)).

In addition, it is worth mentioning that besides simplest off-site surface vortices we found a variety of other asymmetric vortex solitons families. Most representative examples of on-site vortices that reside mainly at the left or at the right of the interface are shown in Figs. 1(d) – 1(f) (notice, that upon searching for such vortices we translated the lattice by $\pi/\Omega$ in the vertical direction for convenience). We found that on-site vortices are much more sensitive to variation in the height of refractive index step $\delta p$ and typically require $\delta p < p$, so that the existence domain substantially departs from that for their off-site counterparts. Nevertheless, on-site vortices also feature strongly asymmetric shapes (see, e.g. Fig. 1(e)), and can be made completely stable in certain domains of their existence. Finally, we would like to remark that we also found that lattices with *defocusing nonlinearity* also can support asymmetric vortex solitons.

We thus conclude stressing that we have reported, for the first time to our knowledge, the existence of surface vortex solitons. We found that such vortex solitons exist only when the refractive index at the surface does not exceed a critical value, dictated by the depth of the lattices, they are stable and robust under proper conditions, and they feature strongly asymmetric and noncanonical nature. Here we addressed the interface between two distinct periodic lattices with focusing nonlinearity, but results have implications for other physical settings. Also, we believe that our findings reported here motivate the search for general topological solitons supported by surfaces.

*On leave from the Physics Department of M. V. Lomonosov Moscow State University, Moscow, Russia. This work has been partially supported by the Ramon-y-Cajal program, by the Government of Spain through grant BFM2002-2861, and by CONACyT under the project 46552.

# Figure captions

Figure 1 (color online). Field modulus distributions for vortex surface solitons at $b=8$, $p=4.5$, $S=0.05$, and (a) $\delta p=1$, (b) $\delta p=4$. (c) Lattice profile at $p=4.5$ and $\delta p=1$. Field modulus distributions for the highly asymmetric vortex solitons at $b=3$, $p=4$, $S=0.2$, and (d) $\delta p=0.8$, (e) $\delta p=1.3$, (f) $\delta p=0.6$. In all cases $\Omega=4$. Vertical dashed lines indicate interface position.

Figure 2. (a) Domains of existence of surface vortex solitons at $(\delta p, b)$ plane. (b) Energy flow versus propagation constant at $p=4.5$. Points marked by circles correspond to solitons shown in Figs. 1(a) and 1(b). (c) Critical value of $\delta p$ versus lattice depth. In all cases $\Omega=4$, $S=0.05$.

Figure 3. Real part of perturbation growth rate versus propagation constant for $p=4.5$ at $\delta p=0.5$ (a) and $\delta p=1$ (b). (c) Stability and instability domains for surface vortex solitons on $(\delta p, b)$ plane at $p=4.5$. In all cases $\Omega=4$, $S=0.05$.

Figure 4 (color online). Propagation dynamics of surface vortex solitons with $b=8$ (a) and $b=6.65$ (b) at $p=4.5$ and $\delta p=4$ in the presence of white input noise with variance $\sigma_{\text{noise}}^2=0.01$. Vertical dashed lines indicate interface position. In all cases $\Omega=4$, $S=0.05$.

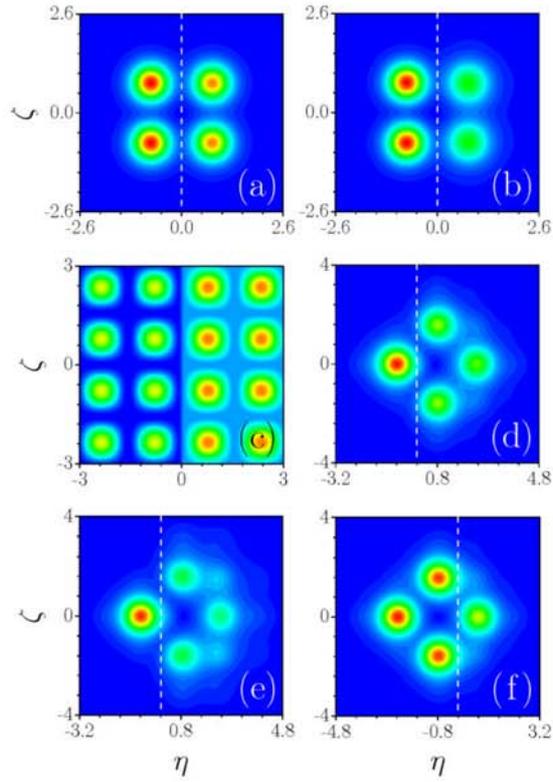

Figure 1 (color online). Field modulus distributions for vortex surface solitons at $b=8$, $p=4.5$, $S=0.05$, and (a) $\delta p=1$, (b) $\delta p=4$. (c) Lattice profile at $p=4.5$ and $\delta p=1$. Field modulus distributions for the highly asymmetric vortex solitons at $b=3$, $p=4$, $S=0.2$, and (d) $\delta p=0.8$, (e) $\delta p=1.3$, (f) $\delta p=0.6$. In all cases $\Omega=4$. Vertical dashed lines indicate interface position.

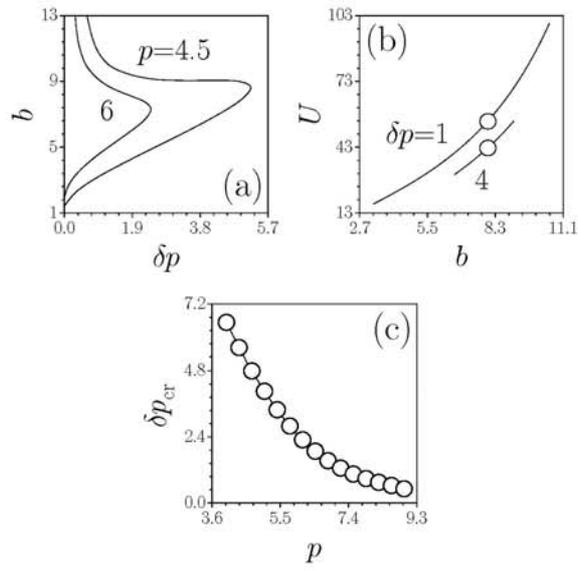

Figure 2. (a) Domains of existence of surface vortex solitons at $(\delta p, b)$ plane. (b) Energy flow versus propagation constant at $p = 4.5$. Points marked by circles correspond to solitons shown in Figs. 1(a) and 1(b). (c) Critical value of $\delta p$ versus lattice depth. In all cases $\Omega = 4$, $S = 0.05$.

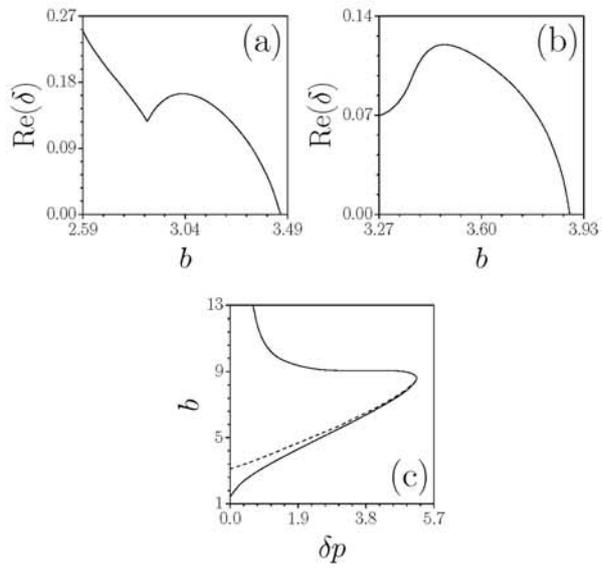

Figure 3. Real part of perturbation growth rate versus propagation constant for $p = 4.5$ at $\delta p = 0.5$ (a) and $\delta p = 1$ (b). (c) Stability and instability domains for surface vortex solitons on $(\delta p, b)$ plane at $p = 4.5$. In all cases $\Omega = 4$, $S = 0.05$.

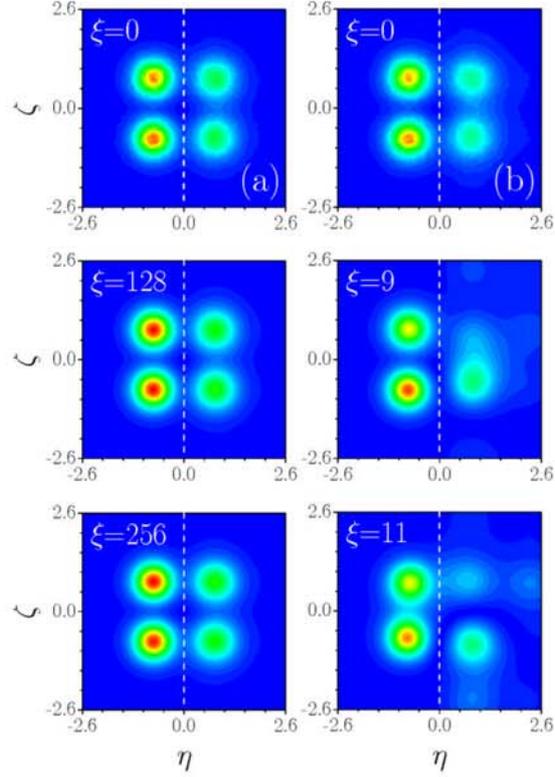

Figure 4 (color online). Propagation dynamics of surface vortex solitons with $b=8$ (a) and $b=6.65$ (b) at $p=4.5$ and $\delta p=4$ in the presence of white input noise with variance $\sigma_{\text{noise}}^2=0.01$. Vertical dashed lines indicate interface position. In all cases $\Omega=4$, $S=0.05$.